# A model of generation of a jet in stratified nonequilibrium plasma

O G Onishchenko [1,2], S N Artekha[2]* and N S Artekha[2,3]

[1] Schmidt Institute of Physics of the Earth of RAS, Moscow 123995, Russia

[2] Space Research Institute of RAS, Moscow 117997, Russia

[3] HSE University, Moscow 105066, Russia

* Corresponding author, E-mail: sergey.arteha@gmail.com

**Abstract:** In the magnetohydrodynamic approximation, system of equations is proposed which analytically describes the initial stage of origination of axially symmetric directed flows in nonequilibrium stratified plasma. The mechanism of generation is based on the Schwarzschild's convective instability and uses the frozen-in flux condition for magnetic field lines. A solution to the nonlinear equation for the stream function is obtained and analyzed, and it is shown that jets with poloidal velocity fields are generated in such a plasma. The corresponding expressions for the R-dependences of the radial and vertical velocity components in the internal and external regions of the jet include Bessel functions and modified Bessel functions. For jets localized in height and radius, the proposed new nonlinear analytical model makes it possible to study their structure and nonlinear dynamics in the radial and vertical directions. The emerging instability in a stratified plasma leads to an increase in the radial and vertical velocities of flows according to the law of the hyperbolic sine. The characteristic growth time depends on the value of the imaginary part of the Brunt–Väisälä frequency. The formation of jets with finite velocity components increasing with time is analyzed. The radial structure of the azimuthal components is determined by the structure of the initial perturbation and can vary with altitude. Along with studying the dynamics of the velocity field, the change in the vertical magnetic field, as well as the dynamics and structure of the emerging toroidal electric current, are investigated.

# 1. Introduction

The object of our research is directed flows in plasma, or jets. These objects are nonstationary vertically elongated structures localized in space (often vortex structures). All of them make an ascending helicoidal motion, the speed and fields of which reach maximum values at some distance from the axis and tend to zero at large distances. A fairly wide class of jets manifests itself in nature. Among them, first of all, we can mention astrophysical jets [1-5], solar coronal jets and loops [6-8]. Along with astrophysical jets, magnetospheric jets are being intensively studied [9-15], as well as jets arising in laboratory experiments [16, 17]. Thus, the study of jets is one of the fundamental problems of physics, since jets cover a wide range of natural phenomena and are also of great interest for practical activities.

We use a magnetohydrodynamic (MHD) approach to describe the emergence of nonrelativistic jets. The full development cycle of a jet can be conditionally divided into the following stages: 1) origin and initial growth, 2) reaching and maintaining of a quasi-stationary state, and 3) attenuation or decay. The damping of jet structures begins with stopping the energy (and mass) influx and occurs due to large viscous dissipation. Currently, there is no universally recognized theory of plasma jet generation. Dynamics of jets is investigated in papers [18-21]. Due to the fact that the suggested models [1-5, 21] have a number of disadvantages, finding new magnetohydrodynamic solutions is an actual task. It seems that the creation of a model of jet generation provides new opportunities for obtaining theoretical and practical results.

The aim of the work is to create a new analytical model of jet generation with several free parameters, which allows us to describe the dynamics of the velocity and magnetic fields of the jet at arbitrary radial distances from its axis. To do this, an exact solution will be obtained in the form of a combination of Bessel functions.

The structure of the article is as follows. The second Section presents a closed system of initial equations of magnetohydrodynamics. The third Section is devoted to the derivation of the basic equation for the stream function describing the developing instability. The fourth Section is devoted to solving the resulting equation in an unstably stratified plasma and creating a new analytical model of jet generation. The fifth Section presents discussion and summary.

**2. Initial equations**

An overview of wave processes in astrophysical plasma is presented in [22]. It is well known that internal gravity waves (IGW) play an important role in the process of mass and energy exchange in stratified media. The frequency range for IGW is limited: $|f| << \omega \leq |\omega_g|$, where $f = 2\Omega \sin\varphi$ is the Coriolis parameter, $\Omega$ is the rotation frequency of a planet or star, and $\varphi$ is latitude. When calculating the growth stage of the jet, we neglect the influence of dissipative processes. The initial system of magnetohydrodynamic equations consists of seven equations: the Euler equation for MHD

$$\frac{d\mathbf{v}}{dt} = -\frac{1}{\rho}\nabla\left(p + \frac{B^2}{2\mu_0}\right) + \frac{(\mathbf{B}\nabla)\mathbf{B}}{\rho\mu_0} + \mathbf{g}, \qquad (1)$$

the equation of the constancy of potential temperature $\theta = p^{1/\gamma_a}/\rho$ along a streamline, which can be written as

$$\frac{d\theta}{dt} = 0, \qquad (2)$$

the incompressibility condition

$$\nabla\cdot(\rho\mathbf{v}) = 0, \qquad (3)$$

two Maxwell's equations

$$\nabla\cdot\mathbf{B} = 0, \qquad (4)$$

$$\nabla\times\mathbf{B} = \mu_0\mathbf{j}, \qquad (5)$$

the condition for the magnetic field to be frozen into the plasma (obtained from the Maxwell equations)

$$\frac{\partial\mathbf{B}}{\partial t} = \nabla\times(\mathbf{v}\times\mathbf{B}), \qquad (6)$$

and to close the entire system of equations, we use the equation of state of an ideal gas

$$\frac{p}{\rho T} = const. \qquad (7)$$

In this system of seven equations, $\mathbf{B}$ is the magnetic field, $\mu_0$ is the magnetic permeability of the vacuum, $\mathbf{j}$ is the electric current density, $\rho$ and $p$ are the density and pressure of the plasma, $\mathbf{v}$ is the velocity of the plasma, $\gamma_a$ is the adiabatic index, $\mathbf{g} = -g\mathbf{e}_z$ is gravitational acceleration on the considered planet or star, $\mathbf{e}_z$ is the unit vector along the vertical, $T$ is the plasma temperature, $d/dt = \partial/\partial t + \mathbf{v}\cdot\nabla$ is the Euler derivative with respect to time. Inside the gradient in Eq. (1), the magnetic field pressure is added to the gas pressure.

## 3. The basic equation for the stream function

We will look for an axially symmetrical solution. We introduce a cylindrical coordinate system $(r, \varphi, z)$ with the $z$ axis in the vertical direction and assume that the values do not depend on the angle $\varphi$: $\partial/\partial\varphi = 0$. Let us write down all physical quantities taking into account weak perturbations:

$$p = p_0(z) + \tilde{p}(t,r,z), \quad \rho = \rho_0(z) + \tilde{\rho}(t,r,z), \quad \mathbf{B} = \mathbf{B}_0 + \tilde{\mathbf{B}}, \qquad (8)$$

where $\mathbf{B}_0 = (0, 0, B_{0z})$ near the magnetic pole (local output); the quantities $p_0$, $\rho_0$, $\mathbf{B}_0$ denote the equilibrium unperturbed values of pressure, density, magnetic field, and $\tilde{p}$, $\tilde{\rho}$ и $\tilde{\mathbf{B}}$ are small perturbations of the corresponding quantities at the initial stage of instability development. In the equilibrium state, two equations follow from Eq. (1): the condition of hydrostatic equilibrium along the z axis

$$\frac{d p_0}{d z} = \rho_0 g, \qquad (9)$$

and the condition for radial dependence

$$\frac{d}{d r}\left(p_0 + \frac{B_0^2}{2\mu_0}\right) = 0. \qquad (10)$$

Equation (9) determines the height dependence of the equilibrium pressure:

$$p_0(z) = p_0 \exp\left(-\frac{z}{H}\right), \quad H = \frac{c_s^2}{\gamma_a g}, \quad c_s = \sqrt{\frac{\gamma_a p_0}{\rho_0}}, \tag{11}$$

where $H$ is the characteristic scale of the height of the plasma medium, and $c_s$ is the speed of sound in the medium.

To obtain the equation for vorticity, we apply the curl operation to both parts of (1):

$$\frac{d\boldsymbol{\omega}}{dt} = (\boldsymbol{\omega}\cdot\nabla)\mathbf{v} + \frac{1}{\rho^2}\left[\nabla\rho\times\nabla\left(p + \frac{B^2}{2\mu_0}\right)\right] + \nabla\times\frac{(\mathbf{B}\nabla)\mathbf{B}}{\rho\mu_0}. \tag{12}$$

We take into account that the values do not depend on $\varphi$: $\partial/\partial\varphi = 0$. In the general case, the velocity of plasma flow $\mathbf{v} = (v_r, v_\varphi, v_z)$ can be decomposed into its poloidal component $\mathbf{v}_p = v_r\hat{\mathbf{e}}_r + v_z\hat{\mathbf{e}}_z$ and azimuthal component $v_\varphi\hat{\mathbf{e}}_\varphi$, i.e. $\mathbf{v} = \mathbf{v}_p + v_\varphi\hat{\mathbf{e}}_\varphi$. Since we have

$$\frac{1}{\rho}\frac{d\rho}{dz} \ll \frac{1}{v_z}\frac{dv_z}{dz}$$

for the real scales of any stream, then as a first approximation, instead of condition (3), we can consider the condition of flow incompressibility $\nabla\cdot\mathbf{v} = 0$ for perturbed motion. In this case, to describe the poloidal motion of the plasma $\mathbf{v}_p = (v_r, 0, v_z)$, the poloidal velocity components can be expressed in terms of the stream function $\psi(t, r, \varphi, z)$:

$$v_r = -\frac{1}{r}\frac{\partial\psi}{\partial z}, \quad v_z = \frac{1}{r}\frac{\partial\psi}{\partial r}. \tag{13}$$

We take into account that on the scale of a real jet

$$\frac{1}{B_{0z}} \frac{dB_{0z}}{dz} \ll 1.$$

Then, in the linear approximation, the last term of Eq. (12) turns into

$$-\frac{1}{\mu_0} \frac{\partial}{\partial r}\left(\frac{B_{0z}}{\rho} \frac{\partial \tilde{B}_z}{\partial z}\right) \mathbf{i}_\varphi \tag{14}$$

and it turns out to be small in comparison with the other terms. In many regions of the magnetosphere of the Earth and planets, as well as in the atmosphere of the Sun and stars, plasma can be considered dense. The validity of neglecting the tension of the magnetic field lines will be checked for this model after all the derivations and calculations at the end of the article. From Eqs. (2), (12) and (13), we obtain the following two equations [23]:

$$\frac{\partial}{\partial t}\left(\Delta^* \psi + \frac{d \ln \rho_0}{dz} \frac{d\psi}{dz}\right) + \frac{1}{r} J(\psi, \Delta^* \psi) = -r \frac{\partial \chi}{\partial r}, \tag{15}$$

$$r \frac{\partial \chi}{\partial t} - \omega_g^2 \frac{\partial \psi}{\partial r} + J(\psi, \chi) = 0, \tag{16}$$

where the Jacobian is

$$J(a,b) = \frac{\partial a}{\partial r} \frac{\partial b}{\partial z} - \frac{\partial a}{\partial z} \frac{\partial b}{\partial r}, \tag{17}$$

for elongated cylindrical structures with $\partial/\partial r \gg \partial/\partial z$, the Grad–Shafranov operator has the form

$$\Delta^* = r \frac{\partial}{\partial r}\left(\frac{1}{r} \frac{\partial}{\partial r}\right), \tag{18}$$

$\chi = \tilde{\rho}/\rho_0$ denotes the normalized density perturbation, and the square of the Brunt–Väisälä frequency is determined by the expression:

$$\omega_g^2 = g\left(\frac{\gamma_a - 1}{\gamma_a H} + \frac{1}{T}\frac{dT}{dz}\right). \tag{19}$$

Suppose, for example, that some lower layer or surface of a star is hotter than the plasma above, then the vertical temperature gradient is negative. Let the magnitude of the second term on the right-hand side of (19) exceed the first term on the right-hand side of (19). A similar situation is described by the well-known Schwarzschild criterion: if in a stratified plasma, the frequency of IGWs is a purely imaginary value $\omega_g^2 < 0$, then their absolute instability is observed. The system of equations (15), (16) can be transformed into one equation for the stream function [23, 24], which is the main one for the further construction of the jet model:

$$\left(\frac{\partial^2}{\partial t^2} + \omega_g^2\right)\Delta^*\psi + \frac{1}{r}\frac{\partial}{\partial t}J(\psi, \Delta^*\psi) = 0. \tag{20}$$

We will consider the case when instability occurs at the time moment $t = 0$, i.e. in (20) we have $\omega_g^2 \to -|\omega_g|^2$. In the opposite case, instability does not occur, and the energy of perturbations is carried away from the region of their occurrence with the help of IGWs.

## 4. A model of jet generation

We will look for a scalar stream function generating poloidal velocity components, in the form

$$\psi(t,r,z) = v_0 r^2 f(z/L) \sinh(\gamma t) \Psi(R), \qquad (21)$$

where $v_0 = const$ is some characteristic poloidal velocity; $\gamma = |\omega_g|$; $R = r/r_0$, $L = const$ is a characteristic spatial vertical scale, such that $L \ll H$; $r_0$ is characteristic radial scale of the structure; $\Psi$ is an unknown function depending on the dimensionless radial distance and $\delta = const$; the function $f(z/L)$ will be determined later. Of course, such a stream function with separating variables is not one-valued, but there are additional regularity conditions for the three components of velocity and pressure. The following boundary conditions must also be met:

1) in the center of the jet: $v_r = 0$ at $r = 0$;

2) at the lower and upper boundaries of the jet: $v_z = 0$ at $z = 0$ and at $z = L$;

3) on the periphery of the vortex: $v_r = v_z = 0$, $\tilde{B}_z = 0$ for $r \to \infty$ ($r \gg r_0$).

We use the method of separation of variables to enable an analytical solution. With such a stream function (21), the main Eq. (20) leads to the following equation

$$J(\psi, \Delta^* \psi) = 0. \qquad (22)$$

The general solution of the nonlinear Eq. (22) can be reduced to solutions of the following linear equation

$$\Delta^* \psi = A\psi, \qquad (23)$$

where the values of *A* are arbitrary constants. The stream function for a real model must be finite and localized in the vertical and radial directions; particularly, the following conditions must be satisfied:

$$\left(\psi, \frac{\partial \psi}{\partial r}\right) \to 0, \qquad (24)$$

when $r \to 0$ and $r \to \infty$. This means the requirements for regularity on the axis of symmetry of the jet and for a sufficiently rapid decrease over large distances. Applying the operator $\Delta^*$ to expression (21), we have:

$$\Delta^* \psi = v_0 f(z/L) \sinh(\gamma t) \left( R^2 \frac{d^2 \Psi}{dR^2} + 3R \frac{d\Psi}{dR} \right). \qquad (25)$$

By choosing $A = \pm \delta^2 / r_0^2$ in Eq. (23) and using (25), we obtain a linear equation for the radial function $\Psi$:

$$R^2 \frac{d^2 \Psi}{dR^2} + 3R \frac{d\Psi}{dR} = \pm \delta^2 R^2 \Psi. \qquad (26)$$

The solution to this equation in the general form can be given using Bessel functions. For a real number $\delta$ and a negative sign on the right, the solutions can be the functions $\Psi(R) = J_1(\delta R)/R$ and $\Psi(R) = Y_1(\delta R)/R$. For a positive sign on the right, the solution can be the functions: $\Psi(R) = I_1(\delta R)/R$ or $\Psi(R) = K_1(\delta R)/R$. For $\delta = 0$ we obtain the solution $\Psi(R) = C_0 + C'/R^2$. Moreover, we see that for any choice, the function $f(z/L)$ can be chosen completely arbitrarily. Note that if the constants on the right-hand side of (26) are various, then the sum of such solutions will no longer be a solution of the original nonlinear Eq. (20), i.e. each solution must have its own scope of applicability. For the internal region of the jet, solutions should lead to zero radial velocity on the axis, and at large distances from the axis, solutions should not oscillate, but rather quickly decrease. Solutions for different areas, including velocity components, must continuously and smoothly transform into each other.

To satisfy conditions (24), we look for a solution to the equation by joining two continuous solutions in the internal region $\psi_{int}(r < r_1)$ and the external region $\psi_{ext}(r > r_1)$. At the boundary of the internal and external regions (at $r = r_1$), the continuity conditions are met:

$$\left(\psi, \frac{\partial \psi}{\partial r}\right)_{int} = \left(\psi, \frac{\partial \psi}{\partial r}\right)_{ext}. \tag{27}$$

To satisfy conditions (24), we look for the function $\Psi$, which is part of the solution (21) of Eq. (20) in the external region in the form:

$$\Psi_{ext}(R) = m \frac{K_1(\delta R)}{R K_1(\delta)}, \tag{28}$$

where the parameters $m$ and $\delta$ are to be determined, and in the internal region we will look for the function $\Psi$ in the form:

$$\Psi_{int}(R) = \frac{J_1(\delta_0 R)}{R J_1(\delta_0)}. \tag{29}$$

Let us first define the characteristic radius $r_0$ ($r_0 < r_1$) as such the distance from the axis, at which the radial velocity reaches an extremum (i.e. from the condition that the derivative is equal to zero). As a result, we have the condition

$$J_0(\delta_0) - J_2(\delta_0) = 0, \tag{30}$$

from which we obtain the value $\delta_0 \approx 1.841184$. Taking into account our chosen solutions (28) and (29), let us rewrite conditions (27) for the separating boundary $r = r_1$ in the form of a system of equations:

$$\begin{cases} \delta_0 K_1(\delta r_1) J_0(\delta_0 r_1) + \delta K_0(\delta r_1) J_1(\delta_0 r_1) = 0, \\ m K_1(\delta r_1) J_1(\delta_0) = K_1(\delta) J_0(\delta_0 r_1). \end{cases} \quad (31)$$

The parameter $\delta$ is arbitrary here. Let us choose $\delta = 2.5$ for definiteness. Then we uniquely find from the first equation of the system (31) that $r_1 \approx 1.737734$; and we calculate from the second equation of (31): $m \approx 3.9363$.

The radial dependence of $\Psi(R)$ and the smoothness of this function on the boundary $R = r_1 / r_0$ (dashed line) are visible from Fig. 1 (here and in what follows, the parameter values we have chosen are used as an example).

Using Eqs. (13), (21) and (28) or (29), we find the radial component of the velocity in the inner $(0 \leq r < r_1)$ and outer $(r_1 \leq r < \infty)$ regions, respectively, as:

$$v_r^{int} = -v_0 \frac{r_0}{L} f'(Z) \sinh(\gamma t) \frac{J_1(\delta_0 R)}{J_1(\delta_0)}, \quad (32)$$

$$v_r^{ext} = -v_0 \frac{r_0}{L} f'(Z) \sinh(\gamma t) m \frac{K_1(\delta R)}{K_1(\delta)}. \quad (33)$$

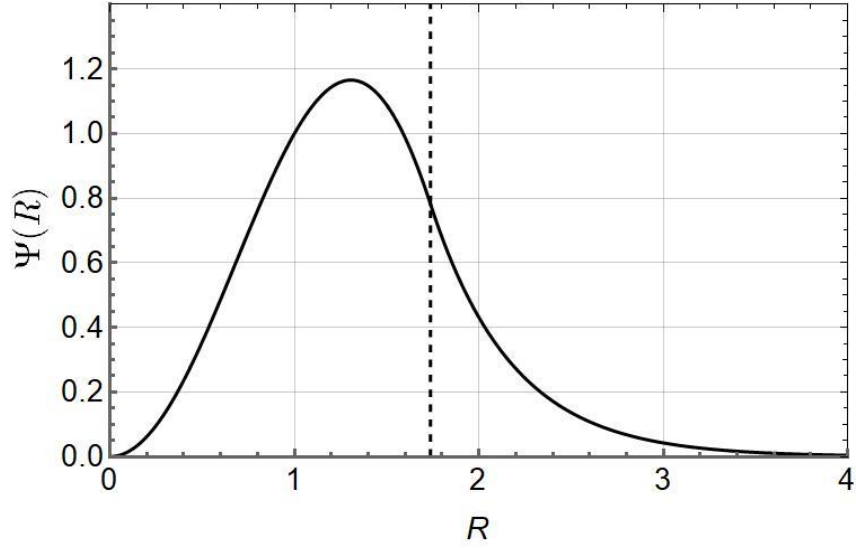

**Fig. 1** Dependence of the function $\Psi(R)$ on the dimensionless radial distance

By choosing the type of function $f(Z)$ of the dimensionless parameter $Z = z/L$, it is possible to achieve different $z$-dependences of the velocity components $v_r$ and $v_z$. For example, one or both components can become zero at the boundaries of the jet, reach a maximum at a certain height or change sign starting from a certain height. For definiteness and simplicity of the graphical representation, we will draw the dependence of the radial velocity component for such a height where $f'(Z) = 1$, and for a height where $f'(Z) = -1$. For example, with the simplest choice

$$f(z/L) = \begin{cases} (z/L), & 0 \leq z \leq L/2; \\ 1-(z/L), & L/2 < z \leq L, \end{cases} \quad (34)$$

we have $f'(Z) = 1$ for $0 \leq z \leq L/2$, and $f'(Z) = -1$ for $L/2 < z \leq L$. The dependence of the $v_r$ component (in units of $v_0$) on the radial distance (for both the inner and outer regions) is shown in Fig. 2 for various values of the $\gamma t$ term and a specific ratio $r_0/L = 0.1$. This shows that the radial component of the velocity and its derivative are continuous throughout the entire region of existence of the jet and decrease sharply at large distances.

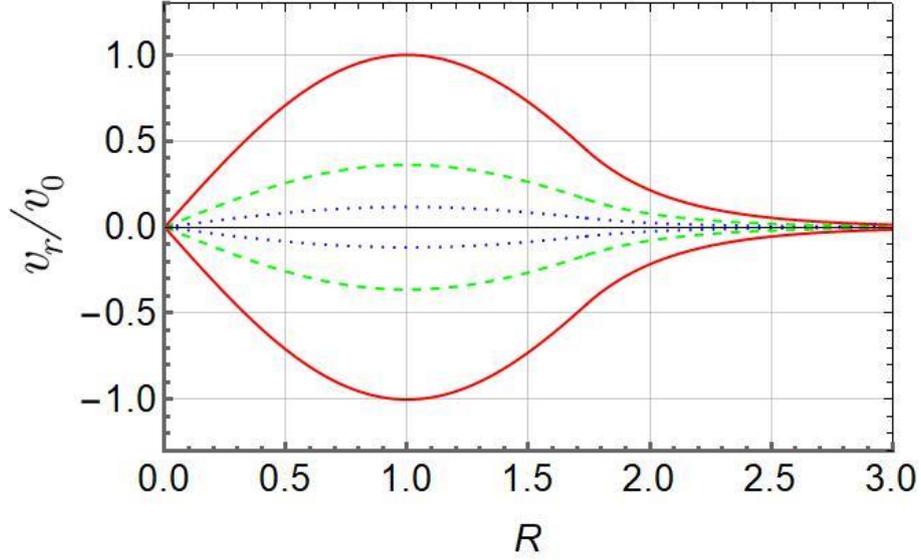

**Fig. 2** Dependence $v_r(R)/v_0$. The dotted, dashed, and solid lines correspond to the values $\gamma t = 1, 2, 3$, respectively; $r_0/L = 0.1$. Negative velocities (inflow) correspond to heights $0 < z < L/2$, positive velocities (outflow) correspond to heights $L/2 < z < L$

By analogy, using Eqs. (13), (21) and (28) or (29), we can obtain expressions for the vertical component of the velocity in the internal region ($0 \leq r < r_1$) and external region ($r_1 \leq r < \infty$), respectively:

$$v_z^{int} = v_0 f(z/L) \sinh(\gamma t) \delta_0 \frac{J_0(\delta_0 R)}{J_1(\delta_0)}, \qquad (35)$$

$$v_z^{ext} = -v_0 f(z/L) \sinh(\gamma t) m\delta \frac{K_0(\delta R)}{K_1(\delta)}, \qquad (36)$$

where the function $f(z/L)$ for the simplest choice is defined in (34). Thus, the poloidal plasma motion arises in convective cells, and similar structure corresponds to vertical jets growing in time. The dependence of the vertical component of velocity (in units of $v_0$) on the dimensionless radial distance $R = r/r_0$ from the axis is shown in Fig. 3 for three values of $\gamma t$.

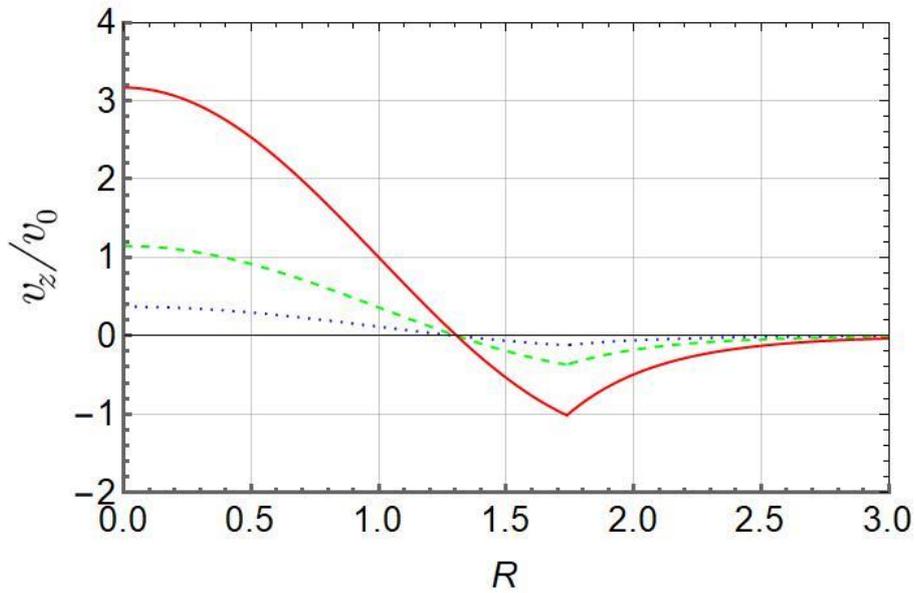

**Fig. 3** Dependence of the dimensionless vertical component of velocity $v_z/v_0$ on the dimensionless radial distance $R = r/r_0$. The picture corresponds to the choice of $f(z/L) = 0.1$. The dotted, dashed and solid lines correspond to $\gamma t = 1, 2, 3$, respectively. If we choose the simplest dependence (34), then the height $z = L/2$ corresponds to the maximum $z$-component of the velocity. At this altitude, the corresponding speeds will be 5 times greater than those shown. Further, the $z$-component of the velocity decreases with height

It is well known that plasma can go into a rotating state for many reasons [25]. Local vortex motions can arise when plasma flows collide, when various inhomogeneities in density, temperature or electromagnetic fields present (symmetry violations, α-effect; Coriolis force; remember the spontaneous appearance of a water funnel in the bathroom). The most natural state of plasma in a magnetic field is a rotating state [26]. To study the development of rotational motion, we write the Euler equation for the azimuthal component of velocity (under the condition $\partial/\partial\varphi = 0$):

$$\frac{\partial v_\varphi}{\partial t} + \frac{v_r}{r}\frac{\partial}{\partial r}(rv_\varphi) + v_z \frac{\partial v_\varphi}{\partial z} = 0, \tag{37}$$

where solutions for the radial and vertical components of velocity can be taken from expressions (32), (33), (35), (36). In order to determine the evolution of the azimuthal velocity component in space and time, we will look for the azimuthal velocity component with the help of the method of separable variables:

$$v_\varphi = v_{\varphi 0} y(t) f_0(z/L) V_{\varphi r}(R), \tag{38}$$

where $v_{\varphi 0} = const$. In order for the solution of Eq. (37) to be determined by such a function with separable variables, we obtain the following system of equations:

$$\frac{d\, y(t)}{d\, t} = \gamma c_0 \sinh(\gamma t)\, y(t), \tag{39}$$

$$\frac{\tilde{v}_r}{Rr_0} + \frac{\tilde{v}_r}{r_0 V_{\varphi r}(R)}\frac{dV_{\varphi r}(R)}{dR} + \frac{\tilde{v}_z}{Lf_0(Z)}\frac{df_0(Z)}{dZ} = -\gamma c_0, \tag{40}$$

where $c_0 = const$ is a certain number (dimensionless constant), and each component with a tilde sign means a part of the factors of the corresponding value without time dependence. Then the solution to Eq. (39) will be the function:

$$y(t) = \exp\{c_0 (\cosh(\gamma t) - 1)\}. \tag{41}$$

Let us now substitute solutions (32), (33), (35), (36) into Eq. (40). As a result, we have:

$$f'(Z)\widehat{V}_r(R) + f'(Z)\frac{R\widehat{V}_r(R)}{V_{\varphi r}(R)}\frac{dV_{\varphi r}(R)}{dR} - \frac{f(Z)\widehat{V}_z(R)}{f_0(Z)}\frac{df_0(Z)}{dZ} = \frac{\gamma c_0 L}{v_0}, \tag{42}$$

where the following designations are introduced for the internal $(0 \leq r < r_1)$ and external $(r_1 \leq r < \infty)$ regions:

$$\widehat{V}_r^{int}(R) = \frac{J_1(\delta_0 R)}{RJ_1(\delta_0)}, \tag{43}$$

$$\widehat{V}_r^{ext}(R) = m\frac{K_1(\delta R)}{RK_1(\delta)}, \tag{44}$$

$$\widehat{V}_z^{int}(R) = \delta_0 \frac{J_0(\delta_0 R)}{J_1(\delta_0)}, \tag{45}$$

$$\widehat{V}_z^{ext}(R) = m\delta \frac{K_0(\delta R)}{K_1(\delta)}. \tag{46}$$

It is easy to see that in the simplest case (34), the choice of solution $f_0(Z) = f(Z)$ leads to a separation of variables. Note that (34) can be changed so that the vertical growth rate $C_1$ for the quantities is arbitrary, and the maximum of the quantities is reached not in the middle, but within the vertical size of the jet at an arbitrary point $Z_1$ of the dimensionless variable $Z$: $0 < Z_1 < 1$. We can select for two areas:

1) $f(Z) = C_1 Z$, $Z \in [0, Z_1]$, and

2) $f(Z) = C_1 Z_1 (1-Z)/(1-Z_1)$, $Z \in [Z_1, 1]$.

In this case, the variables are also separated, but the course of the azimuthal velocity in time changes, and mathematically this leads to a change in the constant $c_0$ on the right side of Eq. (42). Therefore, we will consider the simplest case (34), but we will keep in mind the possibility of a sharp increase in the growth rate of the azimuthal velocity due to a change in the constant $c_0$. Then the solution for the remaining radial function in the inner region ($0 \leq r < r_1$) and outer region ($r_1 \leq r < \infty$) will be:

$$V_{\varphi r}^{int}(R) = \exp\left\{-\int_R^1 \frac{\alpha_{1,2} + V_z^{int}(x) - V_r^{int}(x)}{RV_r^{int}(x)} dx\right\}, \quad (47)$$

$$V_{\varphi r}^{ext}(R) = C\exp\left\{\int_1^R \frac{\alpha_{1,2} + V_z^{ext}(x) - V_r^{ext}(x)}{RV_r^{ext}(x)} dx\right\}, \quad (48)$$

where $C$ is the integration constant for matching two solutions in the internal and external regions; the constants $\alpha_1$ and $\alpha_2$ refer respectively to the lower $0 \leq z \leq L/2$ and upper $L/2 < z \leq L$ parts of the jet height, and for these areas the corresponding substitutions $c_0 = \alpha_1 v_0/(\gamma L)$ and $c_0 = -\alpha_2 v_0/(\gamma L)$ are made. Here the signs are chosen in such a way that at the initial moment of time, the initial azimuthal velocity is continuous in height. If, instead of the minus sign, we choose the plus sign for the upper part of the jet, then this will lead to a very rapid attenuation of the rotation in the upper half of the jet. Different values of the constants $\alpha_1$ and $\alpha_2$ correspond to different differential rotation and different dynamics of vortex motion in height in the regions of radial inflow and outflow. Naturally, for continuity of flow in the horizontal plane, the values of $\alpha_1$ and $\alpha_2$ must be the same in the internal region $(0 \leq r < r_1)$ and the external region $(r_1 \leq r < \infty)$ of the vortex (at the same height). If you choose different values of $\alpha_1^{int}$ and $\alpha_1^{ext}$, then this choice corresponds not just to differential rotation, but to rotation with discontinuity (shift at $r = r_1$). Since we are interested in the case of an unified vortex with continuous azimuthal rotation, then in this case $\alpha_{1,2}^{int} = \alpha_{1,2}^{ext} \equiv \alpha_0$. As a result, we have:

$$c_0 = \frac{\alpha_0 v_0}{\gamma L}. \qquad (49)$$

Thus, setting the five parameters $\delta, \alpha_0, v_0, v_{\varphi 0}$ and $\gamma$ completely determines the structure and dynamics of the jet. We will continue graphical calculations with the previous parameter values by choosing $\alpha_0 = 0.01$. We choose $/v_{\varphi 0}/= v_0$ for definiteness, since the initial background velocities (including the initial rotation) should be comparable. From the viewpoint of physics, this is a question of what initial values are sufficient for the emergence and maintenance of the resulting structure of the jet. In addition to the instability resulting in vertical and radial motion, the rotation is increased and physically maintained due to the conservation of angular momentum. Therefore, the vortex motion will intensify or be maintained only until the entire area with an initial non-zero twist is pulled towards the axis due to the radial inflow (therefore, the value of $\alpha_0$ should be quite small).

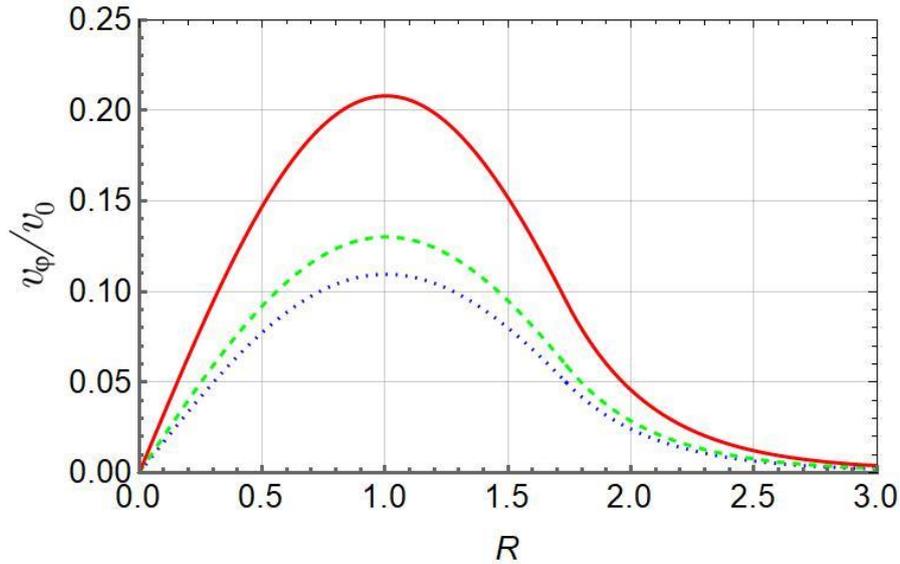

**Fig. 4** Radial dependence $v_\varphi / v_0$ at height $z/L = 0.1$. The dotted line corresponds to the time moment $\gamma t = 3$, the dashed line corresponds to $\gamma t = 4$, and the solid line corresponds to $\gamma t = 5$

The radial localization of the jet can be seen from Figs. 2-4. Thus, Fig. 2 demonstrates the dependence of the normalized radial component of velocity ($v_r / v_0$) on the dimensionless radial distance $R$ (at $r_0 / L = 0.1$) for three different time moments. Radial flows converge on the axis of symmetry of the jet; the radial component of the velocity reaches its maximum absolute value at a distance $R = 1$ from the jet axis. It does not depend on the height up to the height $L/2$, above which it changes sign to the opposite, i.e. the inflow turns into an outflow. If the dependence on $z$ is more complex than (34), then the radial velocity components will differ on opposite sides of a certain dimensionless height $Z_1$ not only in sign, but also in absolute value. At the initial moment, the radial velocity is zero, but over time, the growth of the radial component becomes exponential. Figure 3 shows the dependence of the normalized vertical component of the jet velocity $(v_z/v_0)$ on the same dimensionless distance $R$ (at $z/L = 0.1$) for three different values of $\gamma t$. The increase in the vertical velocity over time obeys Eqs. (35) and (36). The quantity $v_z/v_0$ reaches its maximum value on the axis of the jet. The vertical velocity is directed upwards in the internal area of the jet and downwards in the external area. At a distance $R \approx 1.3$, the vertical component of the velocity becomes zero. The upward flow of the inner region of the jet turns into a downward movement in the region $R > 1.3$ and reaches maximum values at the distance $R = 1.73$. Then the velocity tends to zero at the jet periphery. The vertical velocity reaches its maximum at a height of $z = L/2$ (or $Z_1$ for a more complex dependence on $z$ than (34)).

Figure 4 shows the dependence of the dimensionless azimuthal component of the velocity $v_\varphi/v_0$ on the dimensionless distance $R$. In the case we have chosen, the azimuthal velocity reaches its maximum values at $R = 1$. In the lower half of the vortex, at the maximum in $z$ (for $z = L/2$), the speed will be 5 times greater than that shown in the graph. With an increase in the value of $\gamma t = 1 \to 2 \to 3 \to 4 \to \cdots$ by each unit, the radial and vertical velocities increase approximately *e* times (at first a little more, then the growth tends to an exponential law). At the same time, the growth of the azimuthal velocity in the lower half of the vortex tends to a super-exponential law. Although at first the azimuthal component of the velocity increases slowly due to the smallness of $\alpha_0$, but after $\gamma t = 7$ its growth sharply exceeds the growth of the other two components of the velocity (approximate increase for each unit of $\gamma t = 2 \rightarrow 3 \rightarrow 4 \rightarrow 5 \rightarrow \cdots$: 1.06, 1.19, 1.6, 3.58, 32, 12353, ... times). Therefore, to visually compare the results, we plotted graphs for $\gamma t = 3, 4$ and $5$, rather than depicting this growth at the initial stage, where it is hardly noticeable, or later, where it is huge.

The resulting solution for the jet velocity field and its features, described above, can be represented component by component in the form of contour plots on the $R - Z$ plane (see Fig. 5).

The domain of applicability of the solution in time is limited by the initial stage of jet formation. With respect to spatial variables, the solution is applicable up to distances at which the velocities become background values. Physically, the domain of applicability of the solution is limited by the region where instability exists and the region with non-zero torque. In the case $\omega_g^2 > 0$, instability does not arise, all hyperbolic functions in the solutions transform into the corresponding trigonometric functions, disturbances are carried away from the region of occurrence with the help of IGWs, and the structure does not develop.

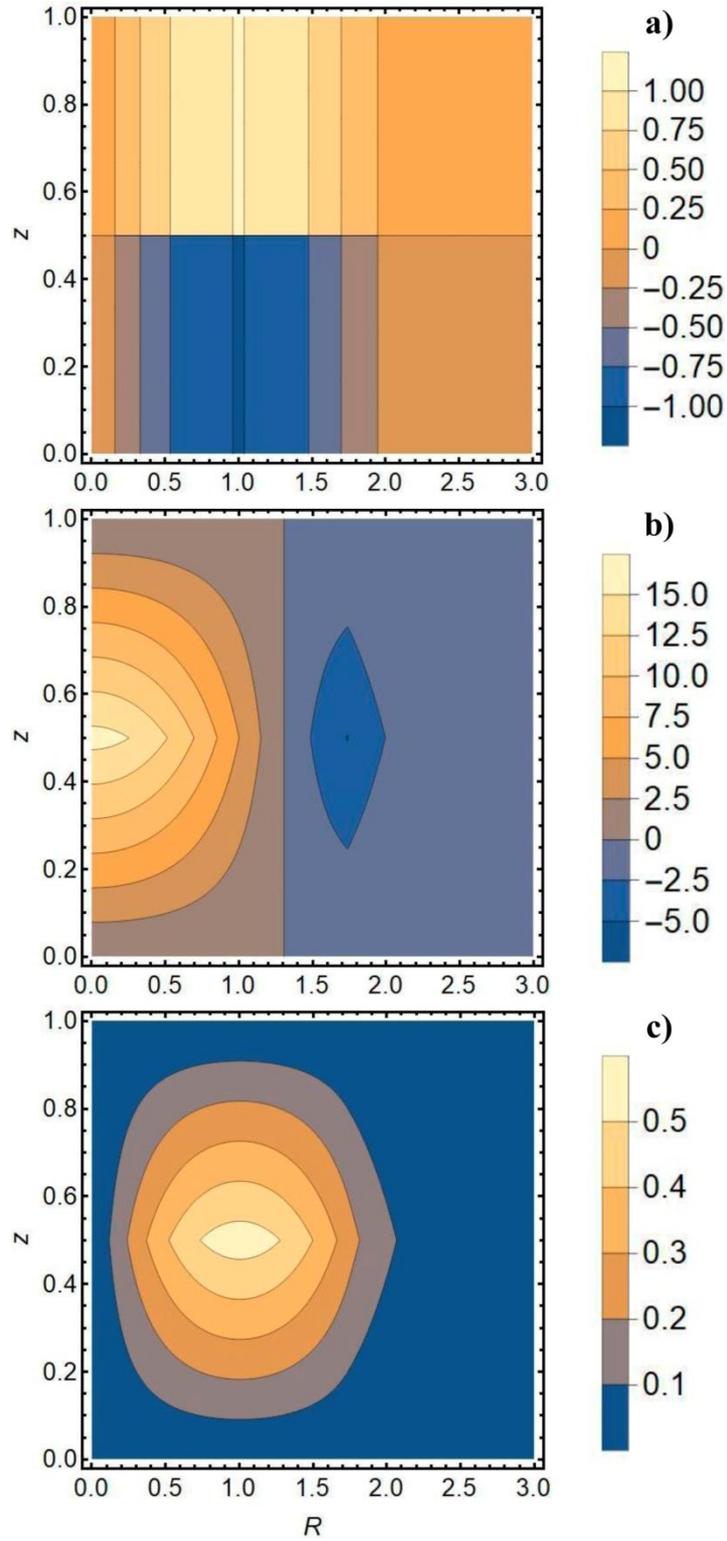

**Fig. 5** The velocity field of the jet for $\gamma t = 3$: a) $v_r/v_0$, b) $v_z/v_0$, c) $v_\varphi/v_0$.

From Eq. (6), expanding the right side, we obtain in a linear approximation:

$$\frac{\partial \tilde{B}_z}{\partial t} = -B_{0z}\left(\frac{\partial v_r}{\partial r} + \frac{v_r}{r}\right). \tag{50}$$

Substituting expressions (35) and (36) for the radial velocity in the internal region and external region of the jet into Eq. (50), we obtain:

$$\tilde{B}_z^{int} = \frac{B_{0z}v_0}{\gamma L} f'(Z)(\cosh(\gamma t) - 1)\delta_0 \frac{J_0(\delta_0 R)}{J_1(\delta_0)}, \tag{51}$$

$$\tilde{B}_z^{ext} = -\frac{B_{0z}v_0}{\gamma L} f'(Z)(\cosh(\gamma t) - 1)m\delta \frac{K_0(\delta R)}{K_1(\delta)}. \tag{52}$$

Now it is easy to see that for this model, the value of (14), obtained from the last term in (12), is equal to zero (because $f'(Z) = const$). Thus, for this model, the tension of the magnetic field lines (the last term in (12)) should be taken into account only if the external magnetic field itself changes noticeably with $z$ on the jet scale. Of course, if we are dealing with a sudden release of some kind of energy (for example, magnetic) in a certain area, and not with convective instability, then it is necessary to build another model taking these factors into account.

Using expressions (51), (52) and Maxwell's Eq. (5), we obtain an expression for the density of the toroidal electric current in the internal and external regions of the jet:

$$j_\varphi^{int} = -\frac{B_{0z}v_0\delta_0^2}{\gamma L \mu_0 r_0} f'(Z)(\cosh(\gamma t) - 1)\frac{J_1(\delta_0 R)}{J_1(\delta_0)}, \tag{53}$$

$$j_\varphi^{ext} = \frac{B_{0z}v_0 m\delta^2}{\gamma L \mu_0 r_0} f'(Z)(\cosh(\gamma t) - 1)\frac{K_1(\delta R)}{K_1(\delta)}. \tag{54}$$

We see that the proposed analytic model of the jet describes the change in the vertical component of the magnetic field and the generation of the associated toroidal electric current. The growth of these quantities tends to be exponential in time; they are localized in the radial direction and change their sign to the opposite one in the external region of the jet.

Thus, the presented jet model is characterized by the following free parameters: characteristic velocity $v_0$, characteristic radial scale of the jet $r_0$ and vertical scale $L$, parameter $\delta$ characterizing the radial structure of the jet, parameter $\alpha_0$ determining the growth rate of the azimuthal velocity, homogeneous external magnetic field $B_0$, increment of convective instability $\gamma$.

## 5. Discussion and conclusions

In hydrodynamics for the Navier – Stokes equations, several analytical solutions are known that describe stationary models of vortices: the Burgers vortex and the Sullivan vortex. However, these dissipative vortices have a number of disadvantages, such as an unlimited growth of values at large distances and the disappearance of the solution (spreading in space) with decreasing viscosity. Therefore, the main goal of this article was the following: in the absence of dissipation, obtain an analytical vortex solution describing a jet localized in the radial and vertical directions, for which the mass conservation condition is satisfied.

In the article, within the framework of ideal MHD, a system of nonlinear equations is derived that describes the behavior of IGWs in a stratified plasma; IGWs in unstable plasma lead to the formation of axially symmetric structures growing over time. Further, the resulting system of Eqs. (15), (16) is equivalent to one simpler Eq. (20), which includes vector nonlinearity. The solutions to this nonlinear equation coincide with the solutions to the linear Eq. (23). As a result, the stream function $\Psi(R)$ is proposed, which makes it possible to reduce the original nonlinear equation to an equation that has various Bessel functions as solutions. By matching the solutions at the boundary separating the inner region of the jet and the outer region of the jet, an analytical solution is found for the radial and vertical components of the velocity, valid for all distances $R$. The radial velocity converges on the axis of symmetry of the jet and reaches its maximum value at a certain radial distance $R$. Starting from a certain height, the inflow is replaced by a radial outflow (see Fig. 5). At the initial moment of time, the radial and vertical velocities are zero, and over time their growth becomes exponential. The vertical flow is greatest on the axis of the jet and reaches its maximum at a certain height (see Fig. 5). The downward movement is realized for the outer area of the structure (see Fig. 5). Thus, the new analytical model helps to describe localized jet structures of poloidal and azimuthal motion (increasing with time) at any radial distances $R$ in plasma. The vortex will rotate differentially. The radial structure of the azimuthal velocity is determined by the structure of the initial disturbance. The azimuthal speed may vary with altitude. The maximum rotation is achieved at a certain height (see Fig. 5). The azimuthal speed can increase according to a super-exponential law. Let us note that vertically, a solution for the jet can be constructed (joined) not only from two elements (with linear dependencies on *z*), but also from a larger number of elements. For example: the base (near-surface layer), the longest central body (if the plasma is stratified and its properties vary greatly, then this body can consist of several parts) and the upper part (scattering region).

Thus, a new model of jet generation in an unstable stratified magnetized plasma is created in the work. This model is created in an axially symmetric approximation using nonlinear equations for internal gravity waves. The jet model is characterized by the following seven free parameters: characteristic velocity $v_0$; characteristic radial scale of the jet $r_0$ and vertical scale *L*; parameter $\delta$ characterizing the radial structure of the jet; parameter $\alpha_0$ determining the rate of growth of the azimuthal velocity; homogeneous external magnetic field $B_0$; increment of convective instability *γ*. It is shown that intense jets are formed very quickly in convectively unstable plasma. The proposed model also describes the change in the vertical component of the magnetic field and the generation of the associated toroidal electric current. It is demonstrated that the growth of the vertical magnetic field and the associated toroidal current tends to exponential in time.

Obviously, the resulting solution relates to a pre-existing instability, when the emerging jet begins to draw energy from the hot near-surface layer (in fact, this is a heat engine). However, in order for this instability to occur, a number of parameters must exceed threshold values. First of all, a superadiabatic temperature gradient should appear in the plasma layer:

$$\frac{dT}{dz} > \frac{\gamma_a - 1}{\gamma_a} \frac{g \rho_0 T}{P_0}.$$

In order for the superadiabatic temperature gradient and the resulting jet to be maintained for some finite time, the surface temperature itself (associated with the excess heat) must be sufficiently high, and such an overheated region must be large enough:

$$\min\{\Delta x_{hot}, \Delta y_{hot}\} \gg 2r_0.$$

In order for the found axisymmetric solution with rotation to be realized, this region of the plasma must have non-zero angular momentum (non-zero helicity). The intensity of the jet and its length $L$ depend on temperature gradients (degree of instability $\gamma$). Also, the jet length can depend on the magnetic field strength [16].

Let us note that changing the equation of state of the plasma (7) will not affect the form of the obtained analytical solution and will only change expression (19) for the Brunt – Väisälä frequency. But in order to describe the saturation stage and determine the limit of instability growth, it is necessary to take into account dissipation (damping within the framework of the Navier – Stokes equation). This complication makes it difficult to find analytical solutions. Various plasma instabilities can also limit the length and lifetime of the jet. For example, azimuthal disturbances can lead to the splitting of a single jet into several parts. In addition, part of the plasma energy can be lost as a result of the development of two-stream instability [27]. All these questions can be the subject of further research, but they require numerical calculation of the complete system of equations, which is beyond the scope of this article.

**References**

1. S Chandrasekhar *Astrophys. J.* **124** 232 (1956).

2. R D Blandford and D G Payne *Mon. Not. R. Astr. Soc.* **199** 883 (1982).

3. R V E Lovelace, C Mehanian, C M Mobarry and M E Sulkanen *Astroph. J. Suppl. Ser.* **62** 1 (1986).

4. A Ferrari *Ann. Rev. Astron. Astrophys* **36** 539 (1998).

5. H L Marshall, B P Miller, D S Davis, E S Perlman, M Wise, C R Canizares and D E Harris *Astrophys. J.* **564** 683 (2002).

6. E Scullion, M D Popescu, D Banerjee, J G Doyle and R Erdélyi *Astroph. J.* **704** 1385 (2009).

7. S Wedemeyer-Böhm, E Scullion, O Steiner, V Rouppe, J de La Cruz Rodriguez, V Fedun and R Erdély *Nature* **486** 505 (2012).

8. N E Raouafi, S Patsourakos, E Pariat, P R Young, A C Sterling, A Savcheva, M Shimojo, F Moreno-Insertis, C R DeVore et al. *Space Sci Rev.* **201** 1 (2016).

9. G M Erickson and R A Wolf *Geophys. Res. Lett.* **7** 897 (1980).

10. C Chen and R Wolf *J. Geoph. Res.* **98** 21409 (1993).

11. J Borovsky, R Elphic and H Funsten *J. Plas. Phys.* **57** 1 (1997).

12. C X Chen and R A Wolf *J.Geophys.Res.* **104(A7)** 14613 (1999).

13. E E Grigorenko, J-A Sauvaud, L C Palin, C Jacquey and L M Zelenyi *J. Geophys. Res.: Space Physics* **119** 6553 (2014).

14. M Stepanova and E E Antonova *J. Geophys. Res.: Space Physics* **120** 3702 (2015).

15. M Palmroth, S Raptis, J Suni, T Karlsson, L Turc, A Johlander, U Ganse, Y Pfau-Kempf, X Blanco-Cano et al. *Annales Geophysicae* **39(2)** 289 (2021).

16. R Safari and F Sohbatzadeh *Indian J. Phys.* **89** 495 (2015).

17. T Wolff, R Foest and H Kersten *Eur. Phys. J. D* **77** 34 (2023).

18. O I Bogoyavlenskij *Physics Lett. A* **276** 257 (2000).

19. O G Onishchenko, O A Pokhotelov, W Horton and V Fedun *Phys. Plasmas* **22** 122901-1–122901-5 (2015).

20. O G Onishchenko, V Fedun, A Smolyakov, W Horton, O A Pokhotelov and G Verth *Phys. Plasmas* **25** 054503 (2018).

21. P M Bellan *Phys. Plasmas* **25** 055601 (2018).

22. M A Fedotova, D A Klimachkov and A S Petrosyan *Plasma Physics Reports* **49** 303 (2023).

23. O G Onishchenko, O A Pokhotelov, W Horton and V Fedun *J. Geophys. Res.: Atmospheres* **121** 7197 (2016).

24. O G Onishchenko, O A Pokhotelov, N M Astaf'eva, W Horton and V N Fedun *Phys. Uspekhi* **63** 683 (2020).

25. B Lehnert *Nuclear Fusion* **11** 485 (1971).

26. S N Arteha *Phys. Plasmas* **3** 2849 (1996).

27. Y Zhou, P M Bellan *Phys. Plasmas* **30** 052101 (2023).